\documentclass[12pt]{iopart}
\usepackage{graphicx}

\begin{document}

\title[Scaling behavior of resistivity in $LaNiO_3$ thin films grown on $SrTiO_3$]{Scaling like behaviour of resistivity observed in $LaNiO_3$ thin films grown on $SrTiO_3$ substrate by pulsed laser deposition} 

\author{S Sergeenkov$^{1,2}$,  L Cichetto Jr$^{2,3,4}$, M Zampieri$^{3}$, E Longo$^{3,4}$ and  F M Ara\'{u}jo-Moreira$^{2}$}

\address{$^{1}$ Departamento de F\'{i}sica, CCEN, Universidade Federal da Para\'{i}ba,  58051-970 Jo$\tilde{a}$o Pessoa, PB, Brazil\\
$^{2}$ Departamento de F\'{i}sica, Universidade Federal
de S$\tilde{a}$o Carlos, 13565-905 S$\tilde{a}$o Carlos, SP,  Brazil\\
$^{3}$ LIEC - Department of Chemistry, Universidade Federal de S$\tilde{a}$o Carlos,  13565-905  S$\tilde{a}$o Carlos, SP, Brazil\\
$^{4}$ Institute of Chemistry, Universidade Estadual Paulista - Unesp, 14801-907 Araraquara, SP, Brazil}

\date{ \today}

\begin{abstract}
We discuss the origin of the temperature dependence of resistivity $\rho$ observed in highly oriented $LaNiO_3$ thin films grown on $SrTiO_3$ substrate by a pulsed laser deposition technique. All the experimental data are found to collapse into a single universal curve  $\rho (T,d) \propto [T/T_{sf}(d)]^{3/2}$ for the entire temperature interval ($20K<T<300K$) with $T_{sf}(d)$ being the onset temperature for triggering a resonant scattering of conduction electrons by spin fluctuations in $LaNiO_3/SrTiO_3$ heterostructure. 
\end{abstract}


\maketitle

\section{Introduction}

Recently, ferroelectric thin films have rekindled discussion about their potential applications in non-volatile random memory and microsensors devices [1-4]. Of special interest are  $LaNiO_3$ (LNO) based materials [5,6] which exhibit properties quite different from the other members of the nickelates family $RNiO_3$ ($R$ being a rare-earth element). Namely, LNO does not undergo a metal-insulator transition (MIT) from paramagnetic  metal to antiferromagnetic  insulator. 
Recall that LNO has the perovskite structure with the pseudo-cubic lattice parameter $a=0.38nm$ and when it is manufactured in the form of thin films  it has a rather good compatibility with oxide substrates typically used for deposition, such as  $SrTiO_3$ (STO) and $LaAlO_3$ (LAO), important for applications in ferroelectric FE-RAM.

Based on the substrate properties and intrinsically induced strain in film/substrate heterostructure, it was found [7-10] that magnetic and transport characteristics of deposited LNO films could be drastically changed. More specifically, a  partial suppression of the charge ordering (responsible for MIT in nickelates) can be achieved by simply changing the film thickness which leads to formation of a principally new magnetic structure, the so-called pure spin-density wave (SDW) material, exhibiting properties of an antiferromagnetic  metal [11-16] (with Neel temperatures $50K<T_N<100K$). Besides, an important influence of both composition and strain on the electrical properties of LNO thin films has been reported [17].

This work reports on the successful preparation, characterization and transport measurements of highly $(l00)$-oriented LNO thin films grown on oriented STO substrates by using pulsed laser deposition (PLD) technique. All the obtained resistivity data (as a function of temperature and film thickness) for three different films are found to collapse into a universal curve, exhibiting a scaling like behaviour dominated by conduction electrons scattering on spin fluctuations (supported by formation of SDW within heterostructure interface) for the entire temperature interval. 

\begin{figure}
\centerline{\includegraphics[width=8.0cm, angle=270]{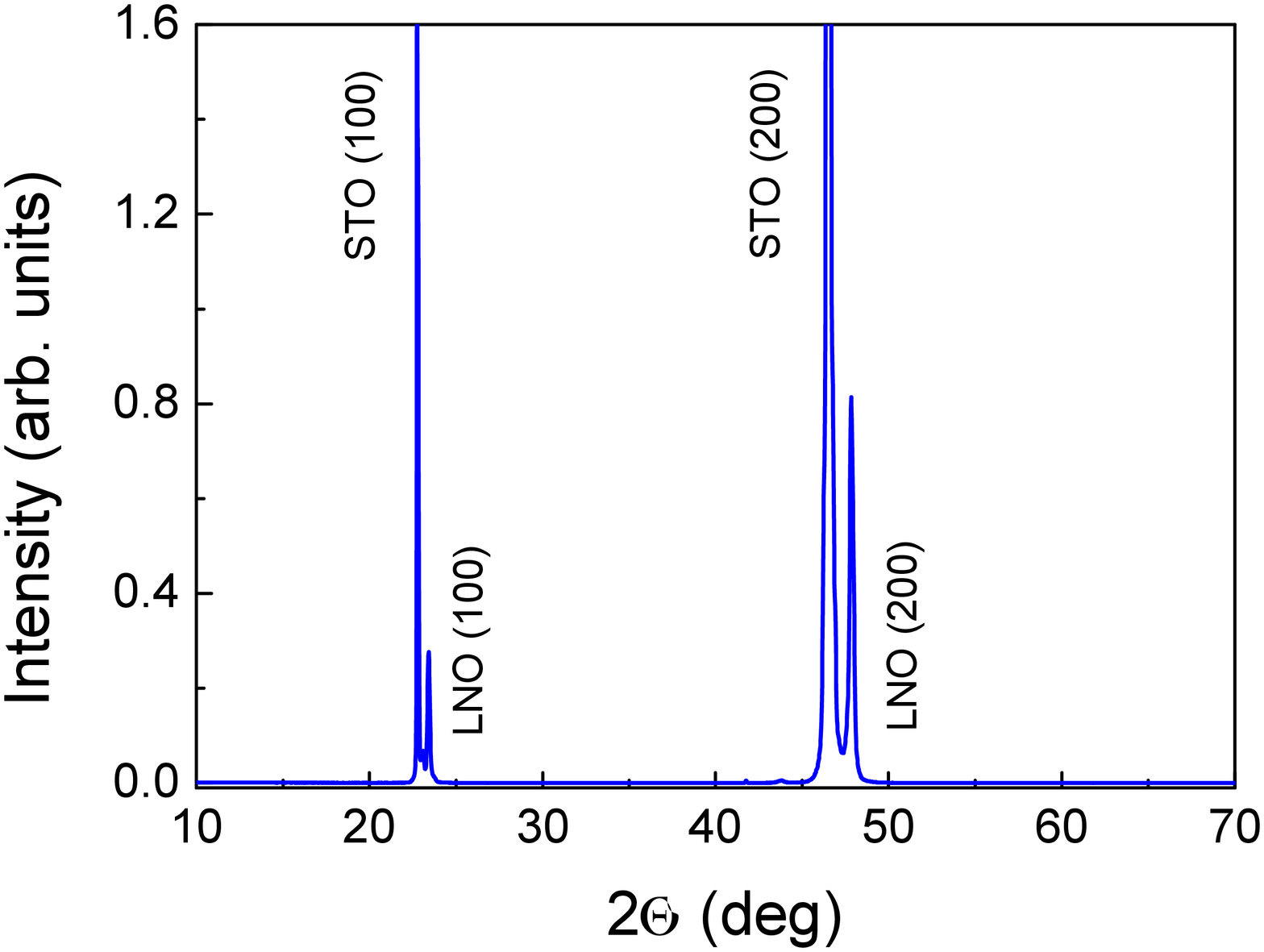}}\vspace{0.5cm}
\caption{Typical XRD spectrum of LNO films deposited on oriented STO substrate. }
\label{fig:fig1}
\end{figure}
\begin{figure}
\centerline{\includegraphics[width=8.50cm]{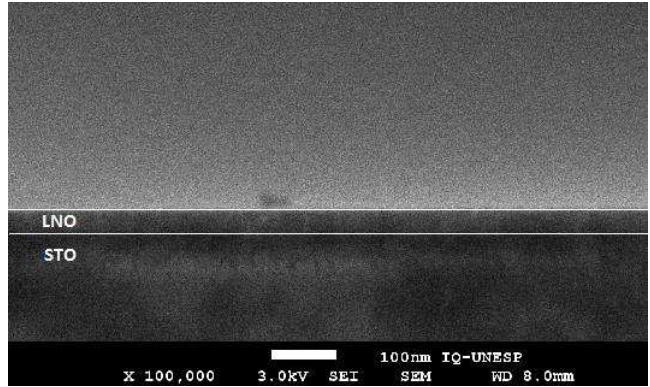}}\vspace{0.5cm}
\caption{Typical FEG SEM image of LNO films deposited on oriented STO substrate. }
\label{fig:fig2}
\end{figure}

\begin{figure}
\centerline{\includegraphics[width=9.50cm]{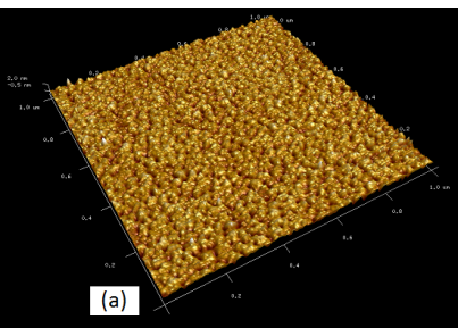}}\vspace{0.25cm}
\centerline{\includegraphics[width=9.50cm]{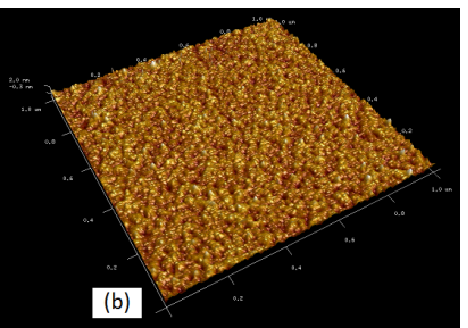}}\vspace{0.25cm}
\centerline{\includegraphics[width=9.50cm]{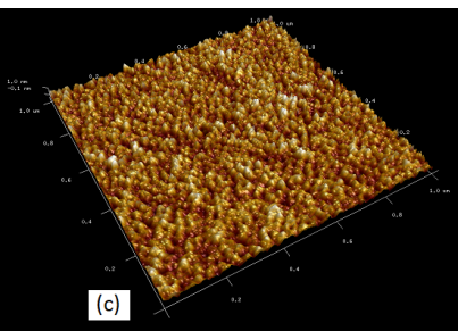}}
\vspace{0.5cm} \caption{AFM surface scans for LNO films of different thickness $d$ deposited on oriented STO substrate: (a) $d=26nm$, (b) $d=33nm$, and (c) $d=46nm$. } 
\label{fig:fig3}
\end{figure}
\begin{figure}
\centerline{\includegraphics[width=7.0cm]{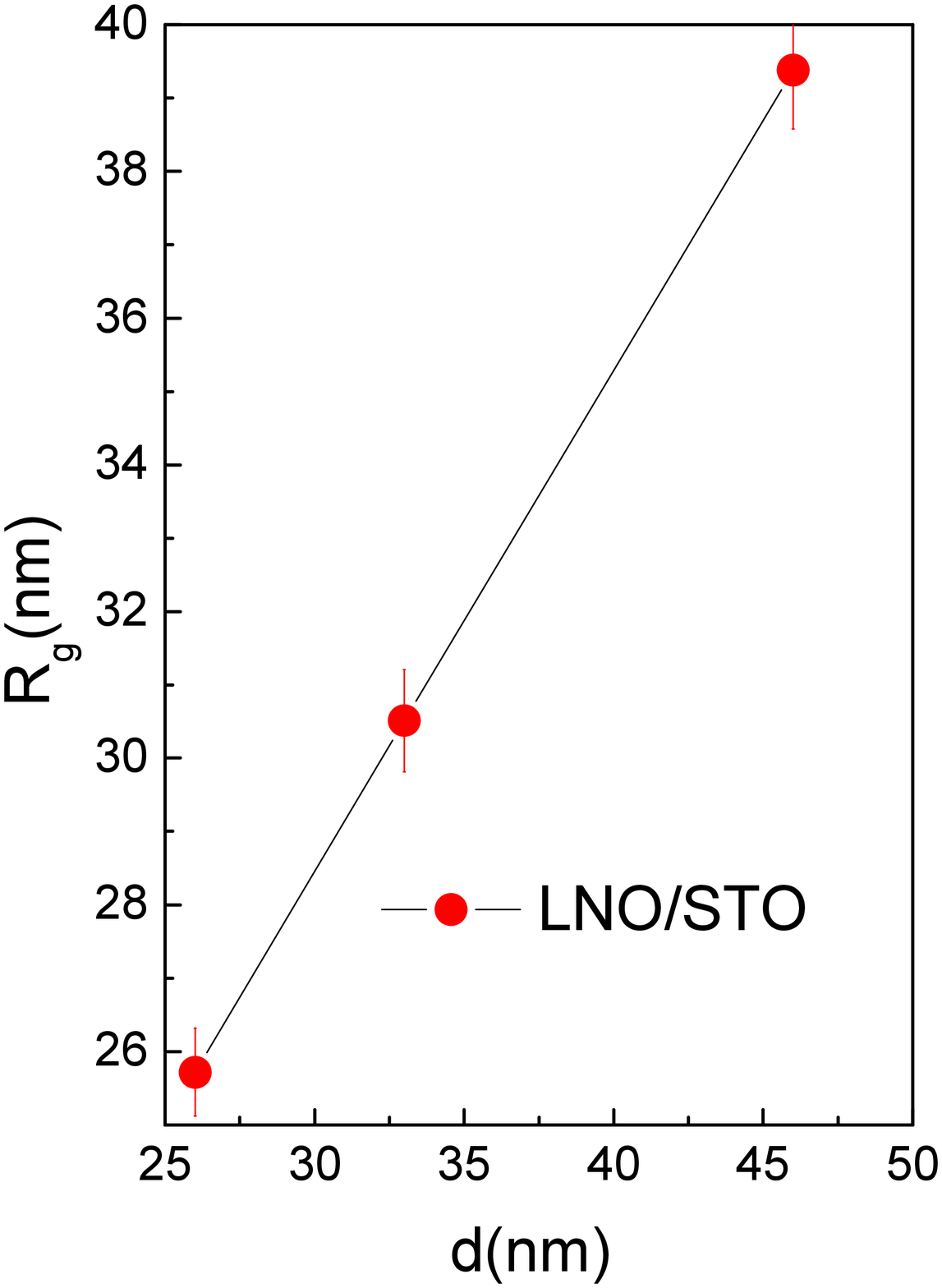}}\vspace{0.5cm}
\caption{Deduced from AFM scans relation between film thickness $d$ and the grain size $R_g$ with error bars.}
\label{fig:fig4}
\end{figure}

\section{Experimental methods}

In order to provide high quality samples (with atomically smooth surfaces compatible with that of the target at a wide range of oxygen pressure), PLD technique was used to deposit thin films of LNO on $(100)$ oriented STO substrate with typical dimensions of $5\times 5 \times 0.5 mm^{3}$. Laser wavelength and repetition rate were $\lambda = 248 nm$ ($KrF$ laser with $25 ns$ pulse duration) and $f = 2 Hz$, respectively.   
The laser energy was maintained constant during the deposition and the beam was focused on the ceramic targets by a quartz lens to a fluency of around $1.2 J cm^{-2}$ for all the samples. During ablation, the target was rotated ($20 rpm$) in order to reduce non uniform erosion and to get the films as homogeneous  as possible.
The heater power was monitored by a computer during the increase and decrease of the temperature. The temperature during deposition was measured by a thermocouple in contact with the heater and the bottom (back side) of the substrate.
The substrate was placed parallel to the target and the distance between them was around $4.8 cm$ which was the best distance found due to the length of the plume.
Before deposition, the base pressure of $P_{base} \le 10^{-7} mbar$ was applied and then the substrates were heated at $800^{\circ}C$ for $20 min$ to get a carbon free and high crystalline surface. Deposition temperature for all films was defined as $T = 625^{\circ}C$ under a flowing oxygen pressure of $P_{dep} = 0.22 mbar$ maintained by a computerized mass flow controller to $80.0$ SCCM (standard cubic centimetres per minute). After the deposition, the samples were \textit{in situ} annealed at the same temperature   for $1h$ under $500 mbar$ oxygen pressure to improve the quality of films and decrease the oxygen vacancy. The dense and crack-free LNO ceramic circular target with diameter of $5cm$ and thickness of $1.25cm$ was prepared from highly pure polymeric precursors by Pechini method [18] using $La_2(CO_3)_3\times H_2O$ ($99.9\%$ Aldrich) and $Ni(OCOCH_3)_2\times 4H_2O$ ($98\%$ Aldrich). Calcination and sintering were performed in the air at $900^{\circ}C$ for $4 h$ and at $1200^{\circ}C$ for $6 h$, respectively. The target was polished after every film deposition to ensure comparable deposition conditions, especially the deposition rate. After that,  the pre-ablation process was carried out for $30s$ to prevent the deposition of the weakly bonded particles. 

The electrical resistivity $\rho (T)$ was measured using the conventional four-probe method. To avoid Joule and Peltier effects, a dc current $I=100\mu A$ was injected (as a one second pulse) successively on both sides of the sample. The voltage drop $V$ across the sample was measured with high accuracy by a $KT256$ nanovoltmeter.

\section{Results and discussion}

\begin{figure}
\centerline{\includegraphics[width=8.0cm]{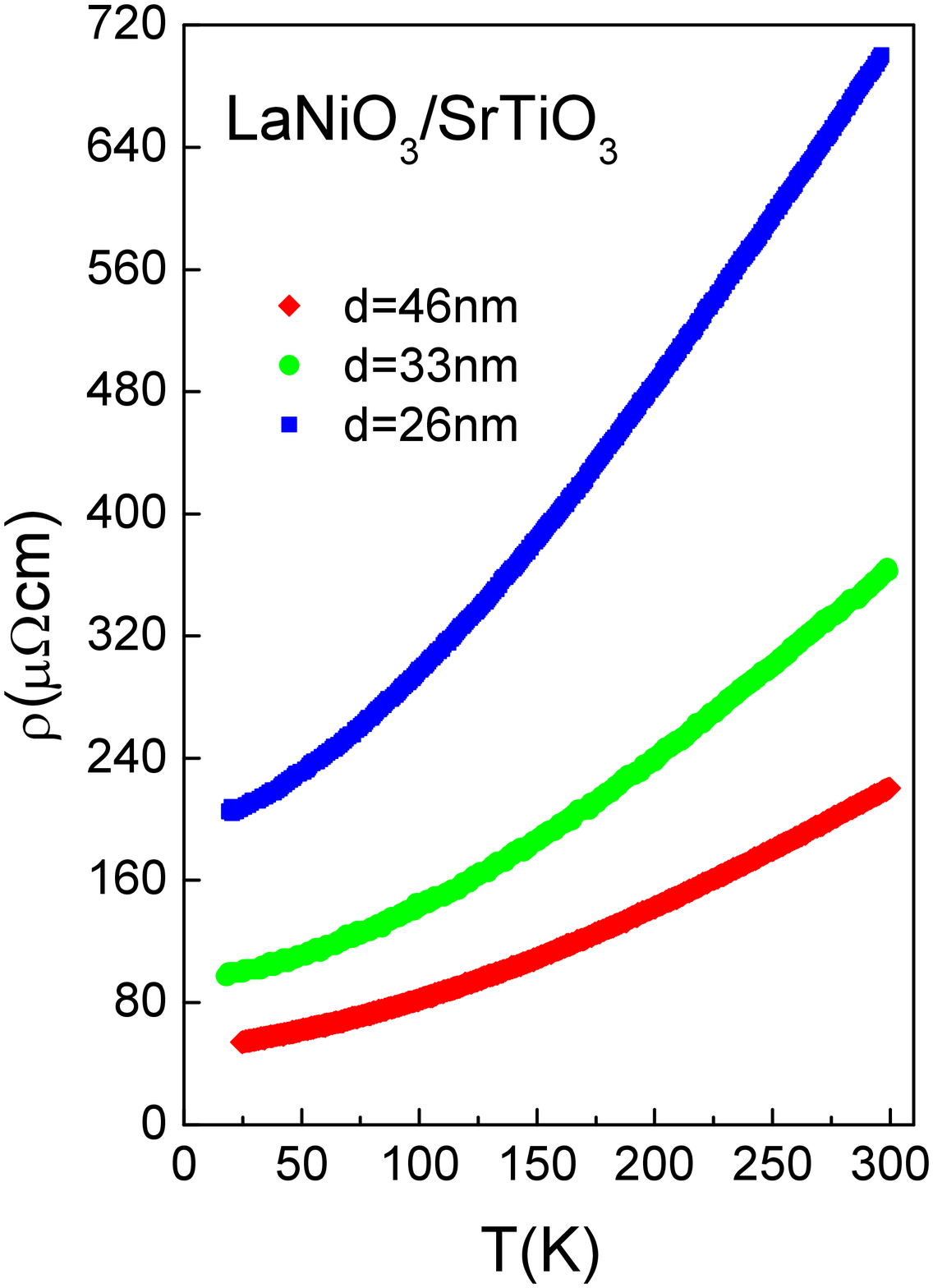}}\vspace{0.5cm}
\caption{Temperature dependence of the resistivity $\rho (T)$ measured for three LNO thin films deposited on oriented STO substrate.}
\label{fig:fig5}
\end{figure}

Microstructure and crystallographic orientation of the films were characterized by X-ray diffraction (XRD) scans. The surface morphology was studied by atomic force microscopy (AFM) and films thickness was confirmed by using field-emission scanning electron microscopy (FEG SEM). Typical XRD spectra and FEG SEM images for the thickest LNO films deposited on STO substrate are shown in Fig.\ref{fig:fig1} and Fig.\ref{fig:fig2}, respectively. AFM scans of  LNO/STO hybrid structures (for three LNO films with thickness of $d=26nm, 33nm$, and $46nm$) are depicted in Fig.\ref{fig:fig3}.  
Fig.\ref{fig:fig4} shows the deduced from AFM scans relation between film thickness $d$ and the grain size $R_g$. The results show that for the thinnest films this relation is practically linear.   Fig.\ref{fig:fig5} shows the typical results for the temperature dependence of the resistivity $\rho (T)$ in our $LaNiO_3/SrTiO_3$  thin films  heterostructure. 

Given the above discussion on appearance of magnetic order in LNO/STO hybrid structure, it is quite reasonable to assume that the observed temperature behaviour of resistivity can be attributed to the manifestation of strong long-range spin fluctuations with a characteristic energy $\hbar \omega _{sf}\simeq k_BT_{sf}$ corresponding to low-energy spin dynamics spectrum measured by inelastic neutron scattering experiments. It should be pointed out that a rather significant scattering of conduction electrons by spin fluctuations is well documented for many different materials, see, e.g., [19-26] and further references therein. Recently, we suggested a simple phenomenological model based on the resonant like features of SDW type spectrum which result in the observed universal temperature dependence of the resistivity, see [27] for more discussion.

\begin{equation}
\rho (T)= \rho_r+\rho_0\left(\frac{T}{T_{sf}}\right)^{3/2}
\end{equation}
where $\rho_r$ is the total residual resistivity and $T_{sf}$ is the onset temperature at which spin fluctuations begin to dominate the scattering process in our thin films  heterostructure.  

\begin{figure}
\centerline{\includegraphics[width=8.0cm]{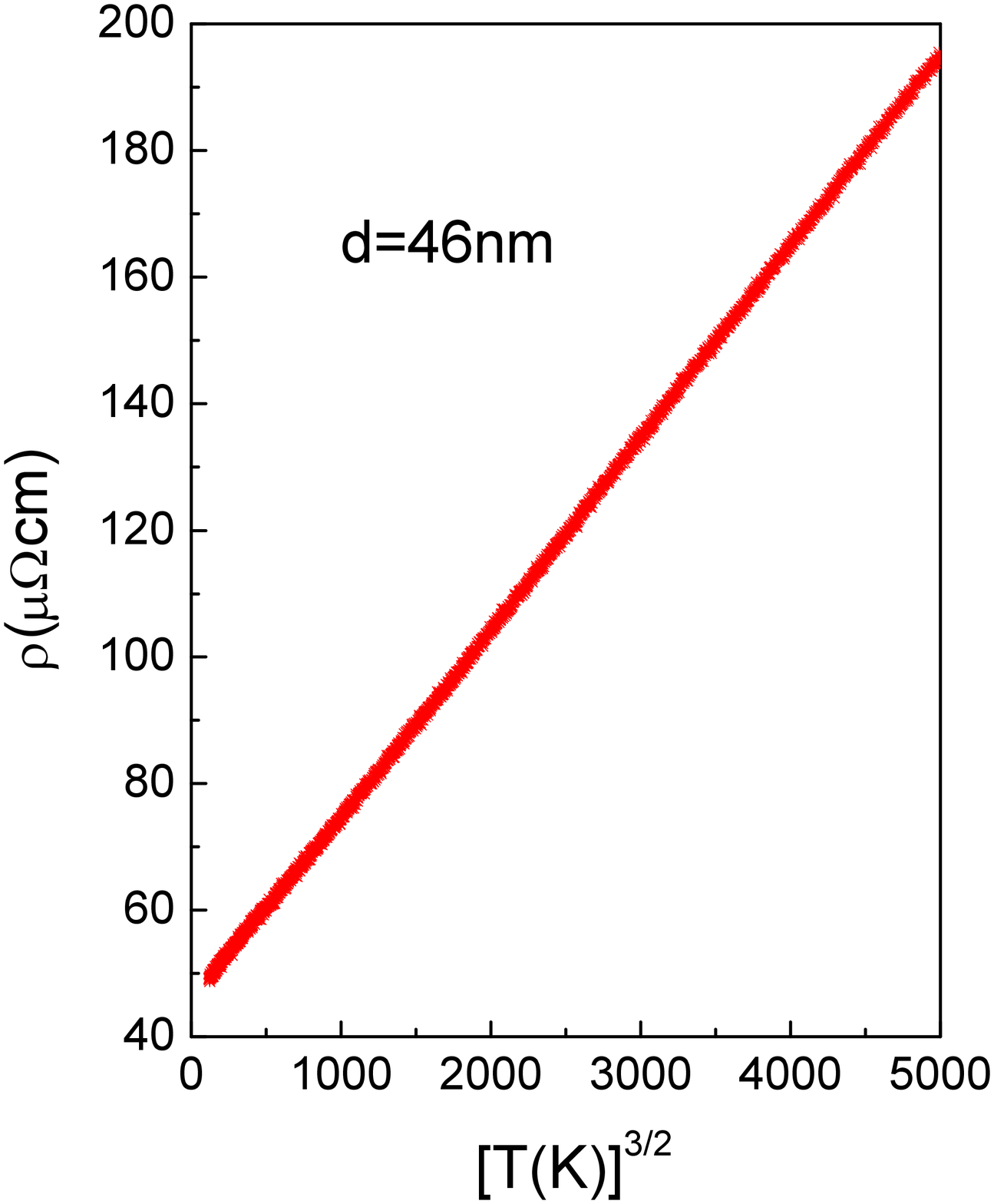}}\vspace{0.5cm}
\caption{ The best fit of the experimental data for the thickest film ($d=46nm$) according to Eq.(1).}
\label{fig:fig6}
\end{figure}

\begin{figure}
\centerline{\includegraphics[width=7.50cm]{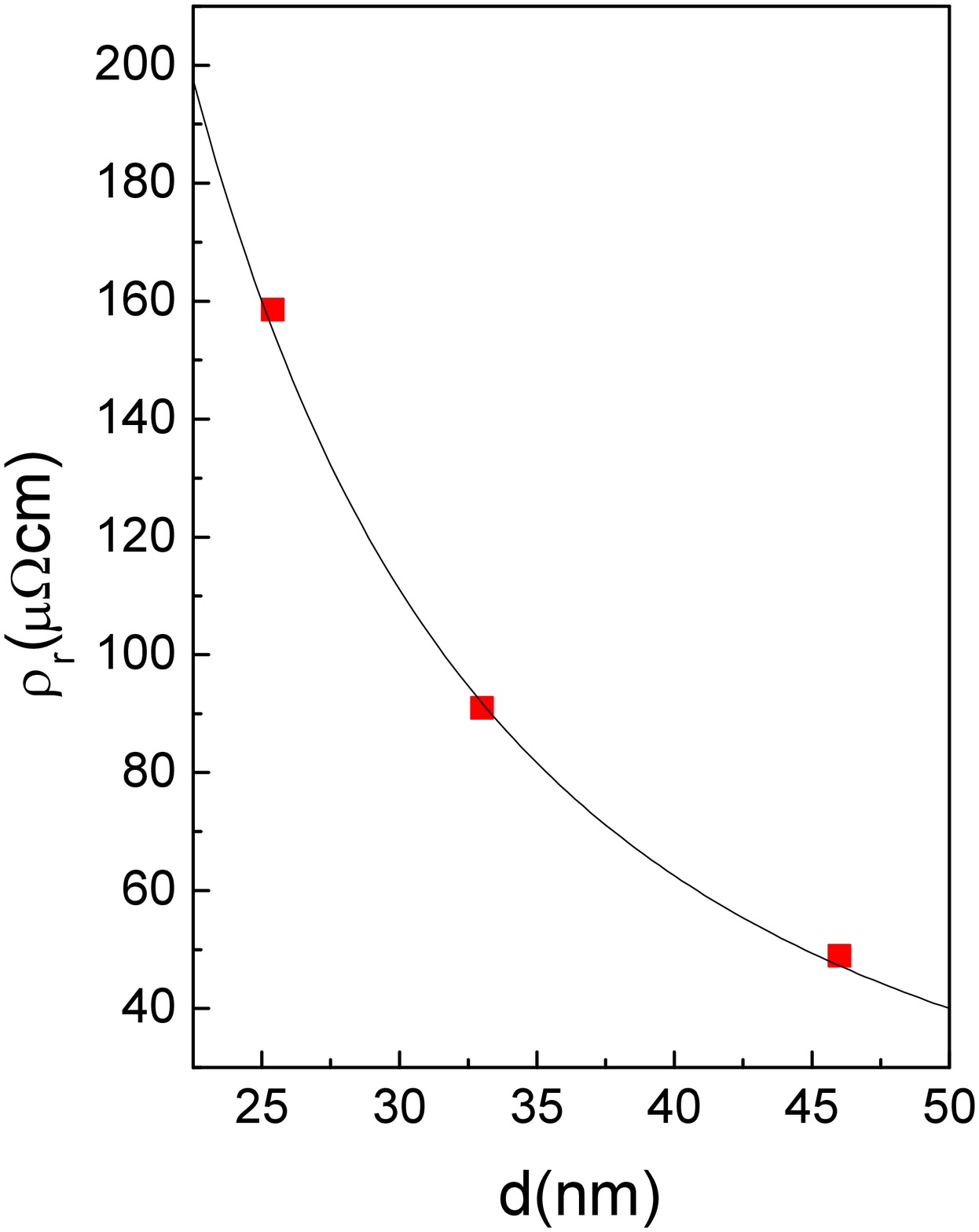}}\vspace{0.5cm}
\caption{Deduced thickness film $d$ dependence of the residual resistivity $\rho_r (d)$ for three LNO thin films deposited on oriented STO.}
\label{fig:fig7}
\end{figure}

\begin{figure}
\centerline{\includegraphics[width=7.50cm]{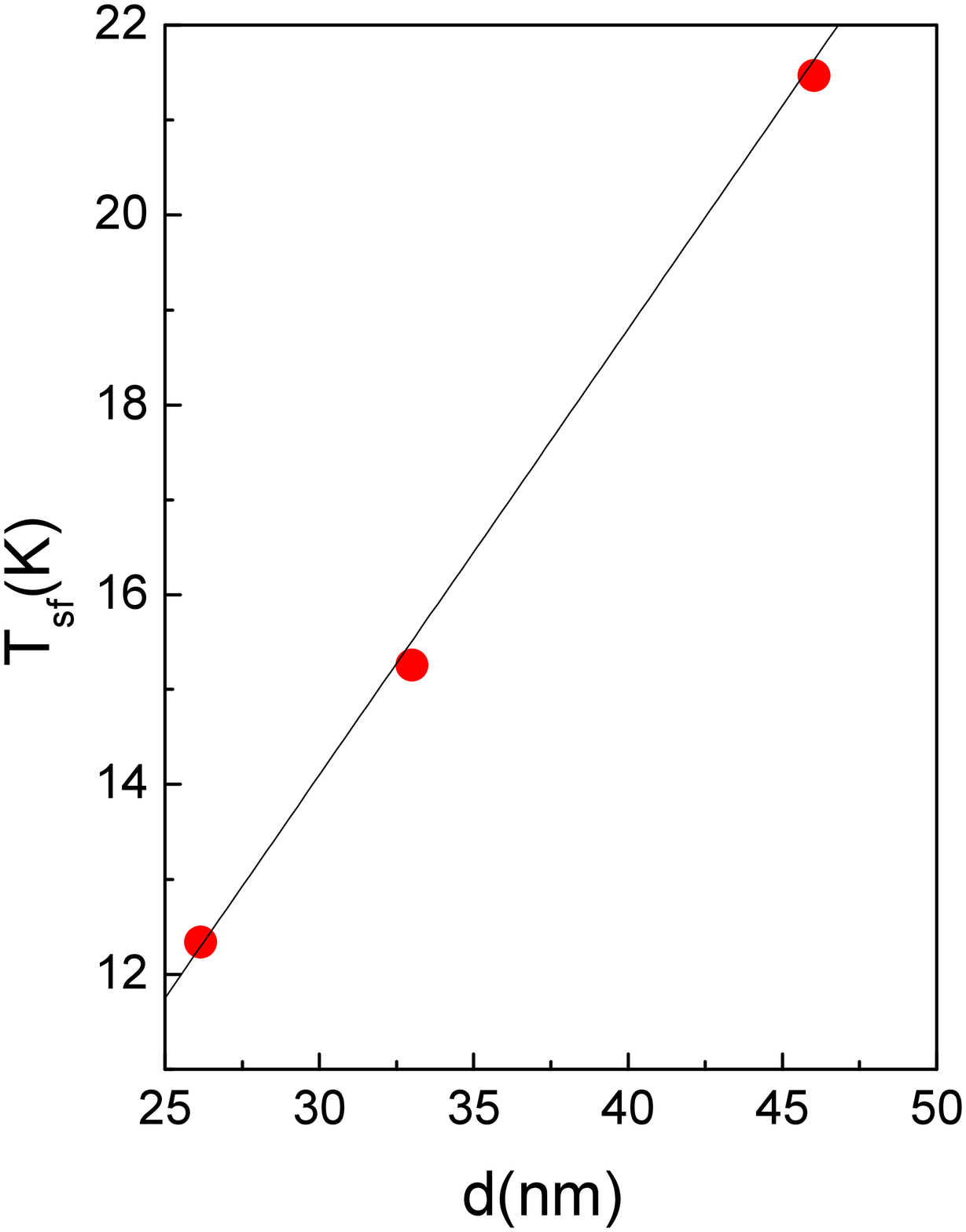}}\vspace{0.5cm}
\caption{Deduced thickness film $d$ dependence of the onset spin-fluctuation temperature  $T_{sf}(d)$ for three LNO thin films deposited on oriented STO.}
\label{fig:fig8}
\end{figure}

Fig.\ref{fig:fig6} shows the best fit of the resistivity data for the thickest film (with $d=46nm$) according to Eq.(1) with $\rho_r=49\mu \Omega cm$, $\rho_0=0.34\mu \Omega cm$ and $T_{sf}=21.5K$. The latter corresponds to $\hbar \omega _{sf}= k_BT_{sf}\simeq 2meV$ in the inelastic neutron scattering spectrum due to antiferromagnetic spin fluctuations [21,27]. 
Now, using equation (1), we can deduce the dependence of the residual resistivity $\rho_r (d)$ and of the onset spin-fluctuation temperature $T_{sf}(d)$ on film thickness $d$. The obtained results are presented in Fig.\ref{fig:fig7} and Fig.\ref{fig:fig8}, respectively. Recall that within the free electron gas model, $\rho_r (d)$ is related to the electron density $n_e$ as $\rho_r (d)\propto 1/n_e(d)$. Thus, expectedly, according to Fig.\ref{fig:fig7}, the electron density increases with increasing the film thickness. On the other hand, according to Fig.\ref{fig:fig8}, the thinner is the film, the lower is the spin fluctuations onset temperature triggering the scattering processes in our films. 
Based on the deduced information about $\rho_r (d)$ and $T_{sf}(d)$, we are able now to fit all the data for all the films by assuming a simple scaling like temperature behaviour of the normalized resistivity $\Delta \rho(T,d)/\rho_r(d)$ as a function of the reduced temperature $T/T_0(d)$ where 

\begin{equation}
\Delta \rho (T,d)= \rho (T,d)-\rho_r(d)
\end{equation}
and 
\begin{equation}
T_0(d)= \left[\frac{\rho_r(d)}{\rho_0}\right]^{2/3}T_{sf}(d)
\end{equation}

\begin{figure}
\centerline{\includegraphics[width=7.50cm]{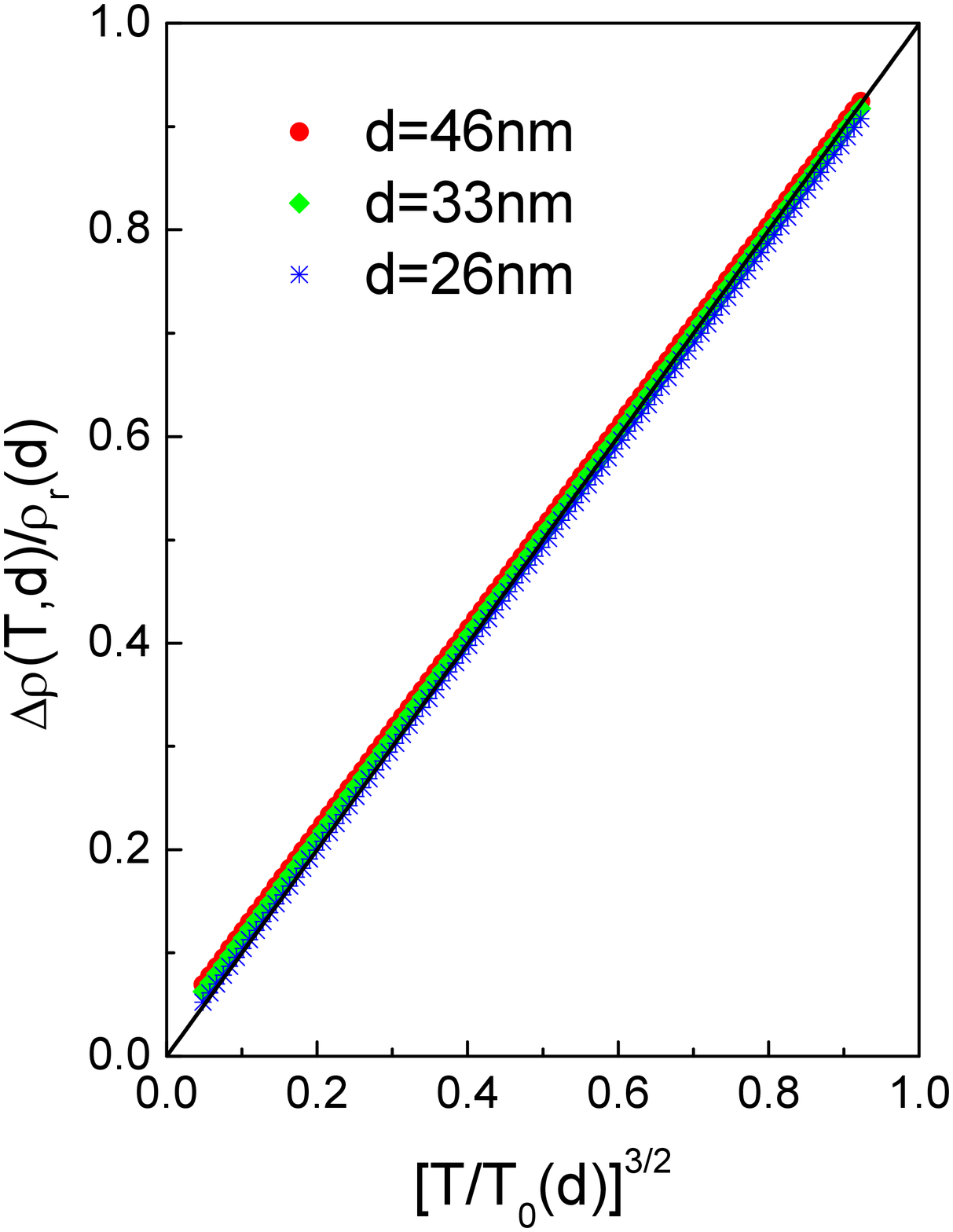}}\vspace{0.5cm}
\caption{ Scaling like behaviour of the normalized resistivity $\Delta \rho(T,d)/\rho_r(d)$ as a function of temperature $T$ and film thickness $d$ for all three films shown in Fig.5. }
\label{fig:fig9}
\end{figure}
Fig.\ref{fig:fig9} is the main result of this paper.  As we can see, all the data points (for all temperatures and all films) collapse nicely into a single line fitted (solid line) by Eqs.(1)-(3). 

And finally, to justify the universality of the observed $T^{3/2}$ behaviour over the entire measured temperatures, it was suggested [20,27] that the resonant scattering of conducting electrons by intraband spin fluctuations replenishes the electron distribution (depleted by interband inelastic scattering) while having a little effect on the current, thus making intraband scattering mechanism responsible for the robustness of the $T^{3/2}$ behaviour in intrinsically strained thin films based heterostructure.

\section{Conclusion}
 
In summary, by analysing the temperature dependence of the resistivity for three different $LaNiO_3$ thin films grown on oriented $SrTiO_3$ substrate (using a pulsed laser  deposition technique) we were able to successfully fit all the experimental data by assuming a universal film thickness dependent scaling like law  dominated by resonant scattering of conducting electrons on spin fluctuations supported by spin-density wave propagation through the interface boundary of $LaNiO_3/SrTiO_3$ heterostructure.

\ack
We are very grateful to H. Kamimura and R.C. Gouveia from NanO LaB for their help with   resistivity measurements. We would like to thank LMA-IQ for allowing us to use FEG-SEM facilities. This work was financially supported by Brazilian agencies FAPESQ (DCR-PB), FAPESP and CNPq. We are very thankful to FAPESP (CEPID CDMF 2013/07296-2 and 2014/01371-5) for continuous support of our project on nickelates.

\end{document}